# Amygdala and insula contributions to dorsal-ventral pathway integration in the prosodic neural network

David I. Leitman[1], J. Christopher Edgar[2], Jeffery Berman[2], Krystal Gamez[1,3],

Sascha Frühholz[4,5,6] and Timothy P.L. Roberts[2]

[1] Department of Psychiatry - Neuropsychiatry Program - Brain Behavior Laboratory, University of Pennsylvania School of Medicine, Philadelphia, PA.
[2] Department of Radiology, Children's Hospital of Philadelphia, Philadelphia, PA.
[3] Department of Graduate Psychology, Immaculata University, Immaculata, PA.
[4]Department of Psychology, University of Zurich, Zurich, Switzerland
[5]Neuroscience Center Zurich, University of Zurich and ETH Zurich, Zurich, Switzerland
[6]Center for Integrative Human Physiology (ZIHP), University of Zurich, Switzerland

*Running title*: Multimodal neuroimaging reveals amygdala contributions to prosody

**Word Count:** Total: 2423; Abstract 150; Introduction: 638; Results: 705; Discussion: 1069. Display Items: 7; Supplemental display Items: 5.

*Submitted to PLOS Biology*

David I. Leitman, Ph.D.

Department of Psychiatry-Neuropsychiatry Program -Brain Behavior Laboratory

University of Pennsylvania

Gates Pavilion 10th floor

3400 Spruce St

Philadelphia, PA 19104-4283

P: (215) 662-7119

F: (215) 662-7903

E: leitman@mail.med.upenn.edu

**ABSTRACT**

Speech prosody enables communication of emotional intentions via modulation of vocal intonations. Reciprocal interactions between superior temporal (STG) and inferior frontal gyri (IFG) have been shown to anchor a neural network for prosodic comprehension, which we refer to as the *Prosody Neural Network* (PNN). Although the amygdala is critical for socio-emotional processing, its integral functional and structural role in processing social information from speech prosody as well as its role in the PNN is largely unexplored including inconsistent recent empirical findings. Here, we used magnetoencephalography and diffusion magnetic resonance imaging of white-matter pathways to establish that the PNN is characterized by (1) a robust amygdala-cortical *functional* connectivity that dynamically evolves as prosodic interpretation progresses, (2) direct *structural* fiber connections between amygdala and STG/IFG traversing a ventral white-matter pathway, and (3) robust amygdala-insula *functional* connectivity and *structural* insula fiber projections to arcuate STG-IFG connections. These findings support a role for functional and structural amygdala-centric ventral pathways in combining speech features to form prosodic percepts. They also highlight insula contributions to prosodic comprehension, potentially via vertical integration of amygdala-centric ventral processing into dorsal pathways responsible for prosodic motor articulation and speech planning.

## INTRODUCTION

When people communicate with one another, along with their exchange of information they also share social and informational cues that guide them into a common emotional understanding (Cherry 1978; Tomasello 2010; Grice 1975). This social phenomenon depends not only on *what* we say, but *how* we say it. Prosody, which is comprised of dynamic acoustic feature modulations, for example, in the pitch, rhythm, and timbre of our vocal intonations, provides a robust channel for sharing social-affective intentions. Like other aspects of language, the listener's interpretation of prosody is the product of distributed cortical and subcortical neural interconnections.

Anatomically, the core neural network of prosodic processing involves interactions between the mid and posterior aspects of the superior and middle temporal gyrus (STG/MTG) and the inferior frontal gyrus (IFG) (S Frühholz and Grandjean 2013; Glasser and Rilling 2008; Thomas Ethofer et al. 2011). During prosodic perception, the STG is believed to extract salient acoustic features from the speech signal and integrate them to form emotional representations (S Frühholz, Ceravolo, and Grandjean 2012; Wiethoff et al. 2008), which are then relayed to the IFG where their meaning and relevance are evaluated (Schirmer and Kotz 2006b; Leitman et al. 2010; Leitman et al. 2011; Wildgruber et al. 2006). Like other auditory signals involving emotion, mental appraisal of prosody is likely to involve prominent input from the amygdala (Sascha Frühholz, Trost, and Kotz 2016). Fear conditioning studies, for example, indicate that amygdala processing may rapidly tag incoming auditory signals to prepare for approach/avoidance responses, and afterwards contribute to more extensive, cortically-centered emotional appraisals (Armony and LeDoux 2010). Other findings indicate that the amygdala is able to decode the emotional meaning from prosody not only when explicitly listing to voices (Fruhholz and Grandjean 2013; S Frühholz, Ceravolo, and Grandjean 2012), but also can also *implicitly* process the emotionality of prosodic signals when emotional voices are presented outside the current focus of attention [e.g. 16, 17].

Despite this recent evidence, it has been surprisingly difficult to reliably demonstrate amygdala activity as well as functional and structural connectivity of the amygdala with other brain regions during prosody processing. First, functional amygdala activity during prosodic perception is inconsistent (S Frühholz, Trost, and Grandjean 2014; Sascha Frühholz, Trost, and Kotz 2016), including p*ositive findings* (Sascha Frühholz, Ceravolo, and Grandjean 2012; Phillips et al. 1998; Morris, Scott, and Dolan 1999; David Sander et al. 2005; Wiethoff et al. 2009; Wildgruber et al. 2006) as well as *null findings* (Grandjean et al. 2005; Mitchell et al. 2004; Kotz et al. 2013; Sascha Frühholz, Trost, and Grandjean 2016; Wiethoff et al. 2008). Second, there seems a paucity of systematic neural network models that incorporate the amygdala (Wildgruber et al. 2009; Sascha Frühholz, Trost, and Kotz 2016), given that no study including functional connectivity analyses yet reported functional connectivity to the PNN (S Frühholz and Grandjean 2012; T Ethofer et al. 2006; Thomas Ethofer et al. 2011). A recent study pointed to a possible functional interconnection of the amygdala to the IFG and not to the STG (S Frühholz et al. 2015), however without defining the structural fiber pathways underlying this functional amygdala-IFG connection. Furthermore, studies including a structural network analysis did yet not report fiber pathways of the amygdala with the PNN (S Frühholz, Gschwind, and Grandjean 2015; Thomas Ethofer et al. 2011), with only one study reporting functional and structural connectivity of the amygdala to the subthalamic nucleus (Peron et al. 2016).

Summarizing these previous findings, first, we have evidence, although partly inconsistent, that the amygdala is sensitive to speech prosody and affectively intonated speech. Second, the amygdala seems functionally connected to the IFG during prosody processing, but evidence for functional connections to other brain regions is missing. Third, the structural fiber connections of the amygdala to the PNN and other relevant brain areas for prosody processing are so far largely undefined. Based on this current evidence, the lack of evidence of functional and structural amygdala connections is surprising, given the importance of the amygdala for socio-emotional processing. In particular, we do not understand how the amygdala influences the STG/MTG and IFG during prosodic processing.

We previously conducted functional magnetic resonance imaging (fMRI) studies in conjunction with a prosody identification paradigm. In these, we not only outlined the PNN network, but also identified a putative role for the amygdala in prosodic processing (Leitman et al. 2010; Leitman et al. 2011). For example, we established that activity in the amygdala and the STG/MTG parametrically increased as signal richness (*i.e.,* prosodic cue salience) became stronger, without being able to determine an amygdala-STG/MTG connectivity. In the IFG, however, *low* prosodic cue salience leading to a high perceptual ambiguity was associated with increased activity, mirrored by higher IFG-STG/MTG functional connectivity. We concluded that increases in cue salience lead to feed-forward feature extraction and integration by the amygdala and STG/MTG, whereas higher IFG activity and IFG-STG/MTG functional connectivity reflect recurrent IFG contributions to the formation of the prosodic percept. In addition to these temporal and frontal cortical regions parametric changes in cue salience also yielded correlated activation changes in a number of additional cortical regions as well and subcortical regions including insula (see *Fig. 1 and Table 1* for a list of task-modulated PNN nodes). Activity within these regions is largely consistent with MRI studies suggesting broadly support these regions as being nodes a PNN, that, also, while lacking direct evidence, suggests amygdala and insula functional integration with cortical regions of the PNN.

| Region | | Description |
|---|---|---|
| STG | ⬆ | Posterior superior temporal gyrus including A1 and planum temporale (PT), roughly Wernicke's area |
| MTG | ⬆ | Posterior middle temporal gyrus including posterior superior temporal sulcus (STS) |
| AMY | ⬆ | Amygdala |
| INS | ⬆ | Insula |
| MOFC | ⬆ | Medial orbitofrontal cortex |
| PCC | ⬆ | Posterior cingulate and precuneus |
| ATG | ⬆ | Anterior superior and middle temporal gyrus and temporal pole |
| SMA | ⬆ | Supplementary motor area. *inconsistent MEG sensitivity to sources in SMA; therefore, SMA was omitted.* |
| TRI | ⬇ | Inferior frontal gyrus pars triangularis (BA 45) |
| OPER | ⬇ | Inferior frontal gyrus pars opicularis (BA 44) |
| ACC | ⬇ | Anterior cingulate gyrus |
| **FMRI activation pattern to cue saliency[5,6]** | | |

Table 1 *Structural-Functional Regions of Interest.* This table indicates the structural regions of interest (ROIs) where prior fMRI studies showed prosody-related task activation which correlated either positively (!) or negatively (") with the presence of emotionally salient acoustic cues (cue saliency). The peak activity within these ROIs served as the basis for our connectivity analysis, as detailed in the methods section. As we noted, SMA was excluded due to inconsistent source localization within this region.

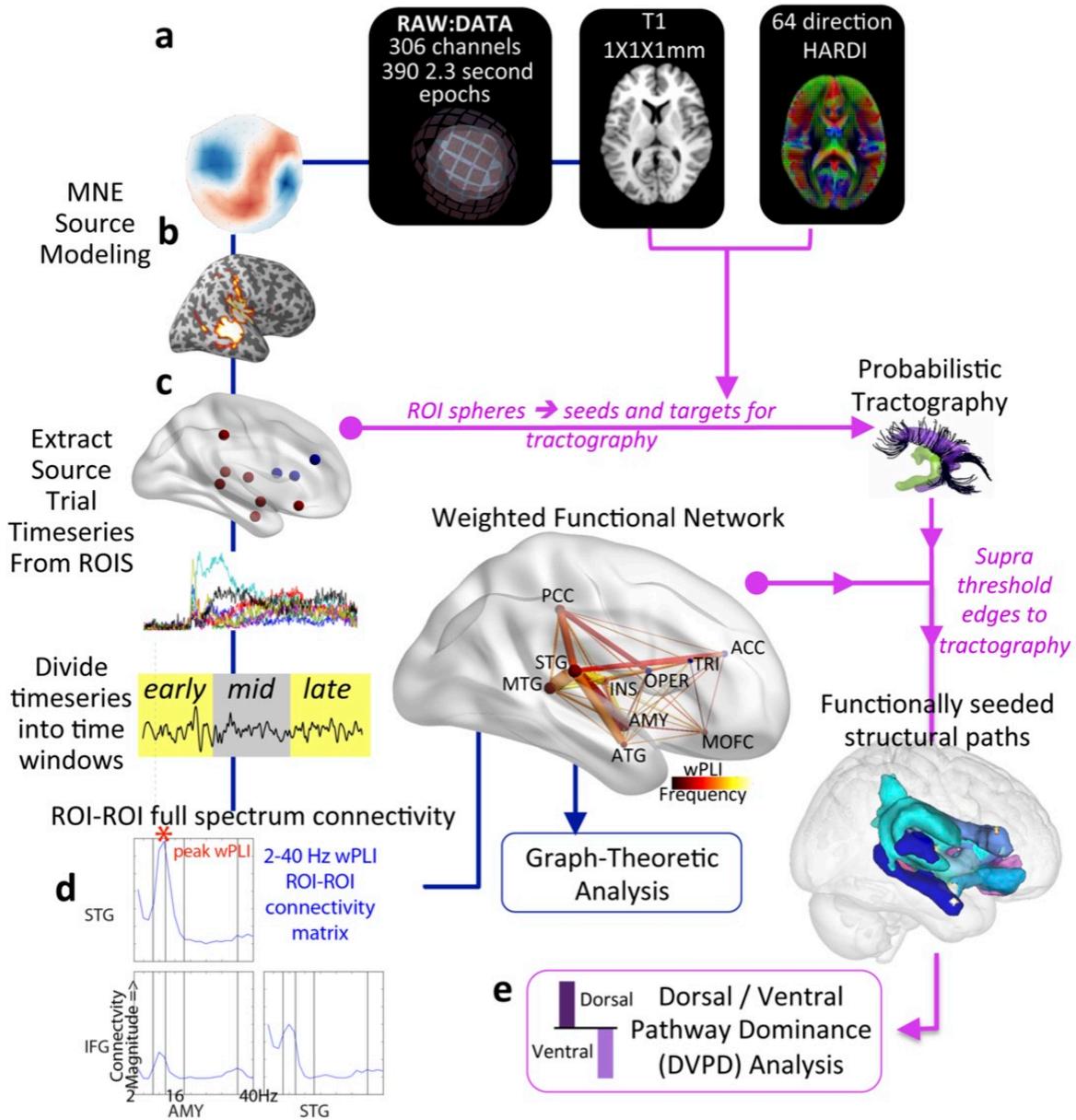

**Figure 1. Flowchart for MEG and MRI analysis.**

Dark blue lines depict functional interactions, and magenta lines reflect structural connectivity. (a) The data from each subject consisted of MEG, high-resolution T1 structural MRI, and B3000 High Angular Resolution Diffusion Imaging (HARDI) Diffusion Weighted Imaging (DWI). (b) Source analysis of MEG data was the result of combining sensor data with cortical and subcortical (hippocampus and amygdala) surface renderings generated from the structural images, using L2 surfaced-based Minimum Norm Estimate (MNE) distributed source model software to perform computations that included the amygdala (AMY). (c) The anatomical ROIs we created were constrained to smaller functional regions that displayed significant fMRI task activation. Within these structural-functional ROIs, a 4-mm sphere was generated around the point of peak MEG activity, as averaged across the different trials (see Table 1 for detailed description). Activities across all virtual magnetic dipoles and orientations passing through this sphere were reduced to single per-trial vectors using Singular Value Decomposition. (d) Time-series data for each of these spherical ROIs were divided into three 600-ms segments and spectrally decomposed into frequency bins ranging from 2-40 Hz. For each time window, the wPLI was calculated between ROI pairs so the overall cross-frequency peak could be identified; the resulting indices formed an adjacency matrix that we used to generate a weighted graphical representation of PNN connectivity. (e) DWI HARDI data were combined with our functional PNN spherical loci to trace connections between supra-threshold wPLI pairs. The resulting tracks were then processed through a waypoint analysis to compare dorsal vs. ventral dominance.

Concerning the PNN, the anatomical path that conveys prosodic processing between the STG/MTG and IFG are so far only sparsely described. One possibility involves the *'dorsal pathway'* that proceeds along the white-matter fiber projections of the superior longitudinal fasciculi and arcuate fasciculi (SLF/AF) (Dick and Tremblay 2012). The relaying of STG/MTG-IFG processing might also occur along a ventral pathway that comprises projections passing subcortically through these hubs via the extreme capsule (EmC) (Saur et al. 2008; Sammler et al. 2015; Sascha Frühholz, Gschwind, and Grandjean 2015; Makris and Pandya 2009). The ventral pathway also has a proximate brain location to the amygdala, and considering this ventral location of the amygdala relative to the PNN (i.e. STG/MTG and IFG), as well as its sensitivity to the emotionality of acoustic prosodic features, we hypothesized that this brain region mainly communicates with STG/MTG (and to a lesser extent IFG) through the subcortical and ventral EmC pathways.

The findings we present here address these questions by using magneto-encephalography (MEG) recordings to measure the temporal evolution of activities within cortical and subcortical processing regions that occur while subjects identified affective-prosodic intent in semantically neutral sentences. Using diffusion MRI mapping, we charted the white-matter pathways that account for these regional activity peaks and thereby created a dynamic functional-structural model of the PNN which establishes that the amygdala (1) is centrally involved in prosodic processing, and (2) communicates with STG/MTG and IFG via a ventral pathway.

**RESULTS**

Participants identified the emotional prosodic tone of semantically neural sentences portraying three primary emotions during MEG recordings of their brain activity (see Experimental Procedures below). Using fMRI-weighted source modeling of our MEG signals, we extracted single-trial time series for all PNN nodes and explored their dynamics using procedures illustrated in *Fig. 1* (see *Supplementary Methods* for further details), For the purposes of examining functional connectivity across both small (40 HZ) and large oscillatory periods (2Hz), we needed time windows that exceeded 0.5s. We therefore divided the period during sentence presentation into three equally-sized 600 millisecond windows that reflect perceptual "chunks" of prosodic information (Pell and Kotz 2011; Fruhholz and Grandjean 2013). This segmentation allowed us to measure the evolution of connectivity changes from early (0-0.6s), middle (0.600-1.2s) and late (1.2-1.8s) time periods post stimulus onset.

*Fig. 2* illustrates event-related activity occurring post stimulus onset during these time periods. Although the timing of activity from PNN sources varied from region to region, it generally peaked within the first 600 milliseconds post-stimulus (M=0.055, SD=0.014), and decreased during the middle (M=0.038, SD=0.0094) and late (M=0.038, SD=0.0095) time windows ($F_{2,304}$=249.97, p<0.0001). Importantly, we found source activity localized to both amygdalae, whose time courses were distinctly different from each other and from sources in insula, which is their closest cortical neighbors suggesting that our MNE surfaced-based joint modeling of cortex and amygdala was successful (see Experimental procedures and *Fig. S1*).

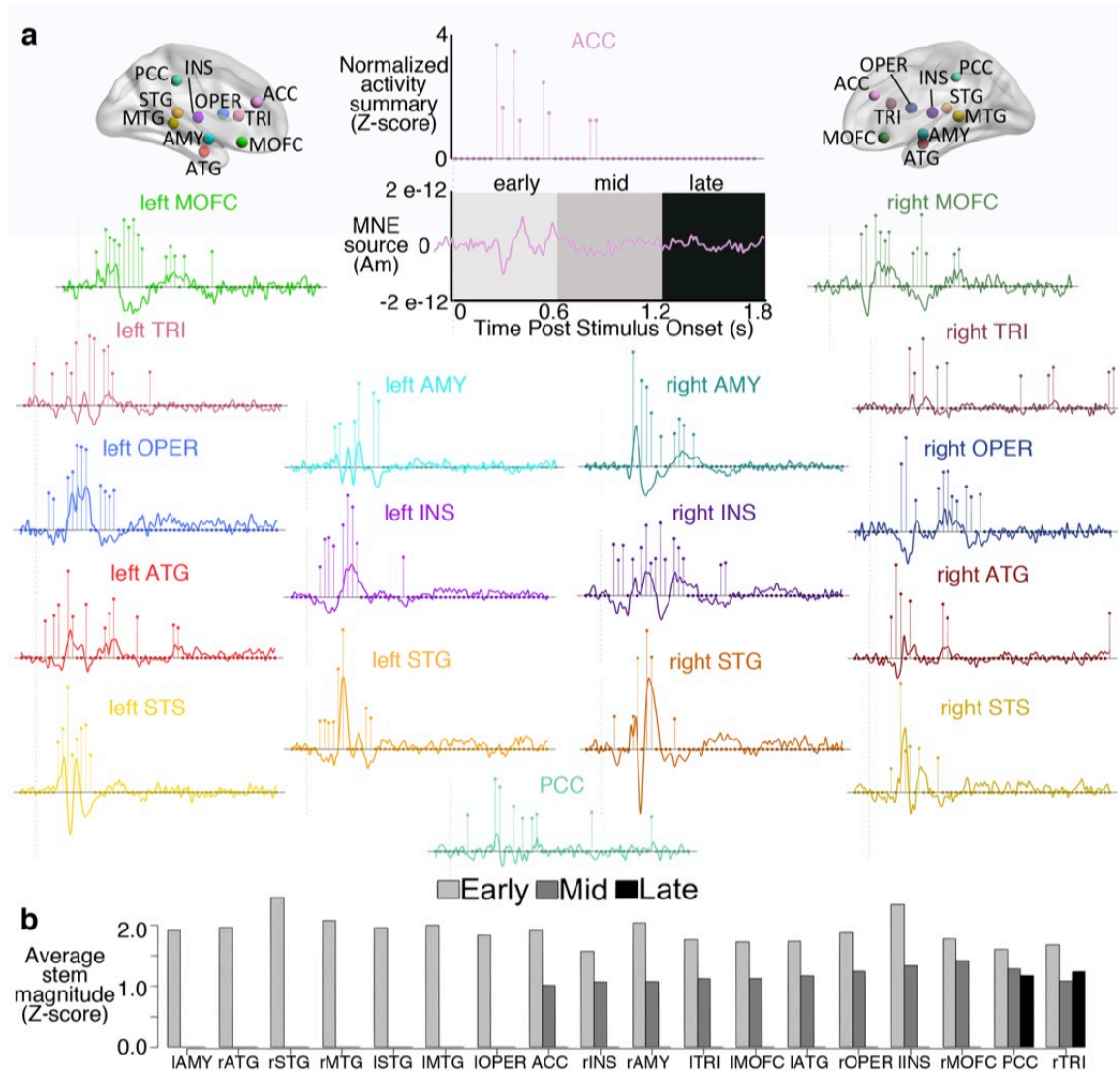

**Figure 1 Prosody neural network time series averaged by region**

(a) The waveforms shown here represent grand-averaged time series across all subjects (N=25) receiving all four stimuli. Waveforms depict source activity magnitude over time, as measured by MNE estimates of dipole source strength in amp-meters (A-m). Stem plots that overlay the grand average waveforms show the ROI-normalized (Z) modulation of activity across time following post-stimulus onset. Stem magnitude is measured in Z units, and each stem averages 20 ms of consecutive time series data. Activity was maximal for all regions during the early (0-0.6 s) window. This was particularly true for the temporal cortical nodes shown in yellow and orange. (b) A summary of normalized activation time courses sorted by time window activity profile for all PNN nodes. Right OPER and insula, ACC, MOFC, TRI, and PCC all displayed sustained evoked activity from 0.6 to 1.2 seconds. In right TRI and PCC, this activity continued to be robust from 1.2 to 1.8 seconds post-stimulus onset. Abbreviations: ACC: Anterior Cingulate Cortex; AMY: Amygdala; ATG Anterior Temporal Gyrus; INS: Insula; MOFC: Medial Orbitofrontal Cortex; OPER: Inferior Frontal Gyrus – Opercularis; MTG: Middle Temporal Gyrus; PCC: Posterior Cingulate Cortex/Precuneus; STG: Superior Temporal Gyrus; TRI: Inferior Frontal Gyrus – Triangularis.

*Functional connectivity across fMRI-weighted MEG time series*

We aimed at examining connectivity patterns within the network of brain regions involved in prosody processing and especially characterize amygdala contributions to this network. Therefore, we used a weighted estimate of the instantaneous Phase Lag Index (wPLI) to determine the magnitude of node-to-node neural connections, which enabled us to construct a weighted-graph model of PNN functional interactions that occur during affective prosodic identification. This model included changes that arise across the early, middle, and late periods following stimulus onset across a broad spectrum of time scales (2-40 Hz). Timescales over 40HZ were not examined given that the average distance between nodes theoretically precluded the possibilities of connectivity at these scales nodes (Feinstein et al. 2011).

    *Fig. 3* depicts connectivity patterns between PNN nodes that occur across these time windows. Overall, the wPLI was maximal for all node-node pairs (*edges*) within high alpha (9-16 Hz) and low beta (16-30 Hz) frequency bands (overall mean peak frequency=16.9 ± 2.8 Hz; see *Tab. S2* for peak wPLI values within specific frequency bands). Peak oscillation frequency increased from the early time window (M=14.6±2.6) to the middle (M=18.0±2.2) and late (M=17.9± 2.2; $F_{2,304}$=106.14; p<0.0001) periods. No differences in connection oscillation frequencies were observed across hemispheres ($t_{262}$=0.727, p=0.47), but connective pairs comprising subcortical nodes (either amygdala and insula) displayed slightly lower oscillation frequencies then cortico-cortico connections overall ($t_{156}$=2.29, p<0.03; Cohen's d=0.13). The reproducibility of connectivity patterns across subjects is illustrated in *Fig. S2, were we illustrate* the homogeneity of high-strength (i.e. high wPLI value) connections across subjects, and across iteratively larger groups of connections using a "Concordance at the Top" (CAT)

approach.

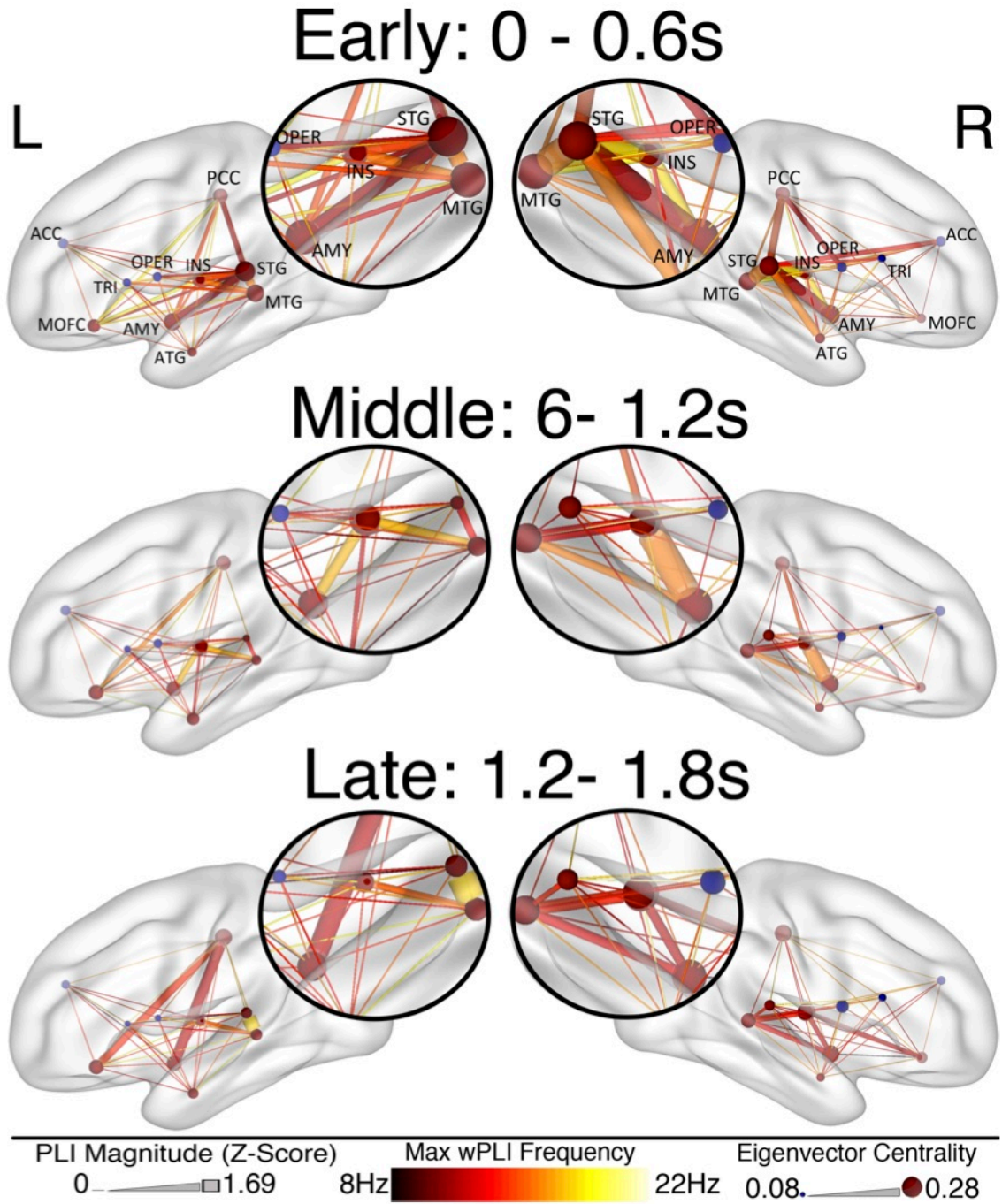

**Figure 2 FMRI-weighted MEG time series showing functional connectivity**

Within-hemisphere weighted connectivity patterns measured across three post-stimulus time windows. Magnified insets centered on AMY-INS-STG-MTG highlight their sustained connectivity patterns across time windows. Blue nodes indicate regions that have higher activity when sparse acoustic features render prosodic intent ambiguous, and red nodes identify areas where activity parallels increased prosodic signal richness. The color of the connecting edges identifies their frequencies, while the relative thickness of these edges reflects their normalized wPLI magnitude; group Z wPLI edge values less than 0.5 are shown as uniform-thickness threads to emphasize high-value edges of interest. During the early time window (at left), bilateral connectivity mainly consisted of interconnections between AMY, INS, and STG. Middle and late time windows in the right hemisphere were characterized by increases in: a) the involvement of PCC and MOFC, b) the connectivity between PCC and MOFC, and c) interactions between MTG and MOFC. Finally, in the left hemisphere, temporal cortical and subcortical connections to frontal control regions (TRI, IFG, and ACC) were maximal during the early time window, then diminished during the middle and late periods. PCC connections to left AMY and MOFC noticeable during the early window become more robust during the middle and late periods. In the right hemisphere, this pattern was not observed; however, temporal, insular, and subcortical regions maintained connections to each other, MOFC, and (to a lesser extent) IFG control regions during the middle and late time windows.

In *Fig. 4,* we then applied Graph theoretic metrics, to these wPLI indices to quantify the robustness of interactions between the amygdala and PNN nodes. We found that amygdala connections *exceeded* those of the collective average PNN nodes in terms of *degree*-number of nodal (regions) connections, *strength*- the overall wPL magnitude of these connection*s,*, *cluster-coefficient*-the proportion of other nodes that have connections with this (amygdala*)* node, and *eigenvalue centrality*- an estimate of centrality based on this nodes connections with other highly connected nodes,. These metric demonstrate amygdala's centrality and 'hub' properties the within the PNN.

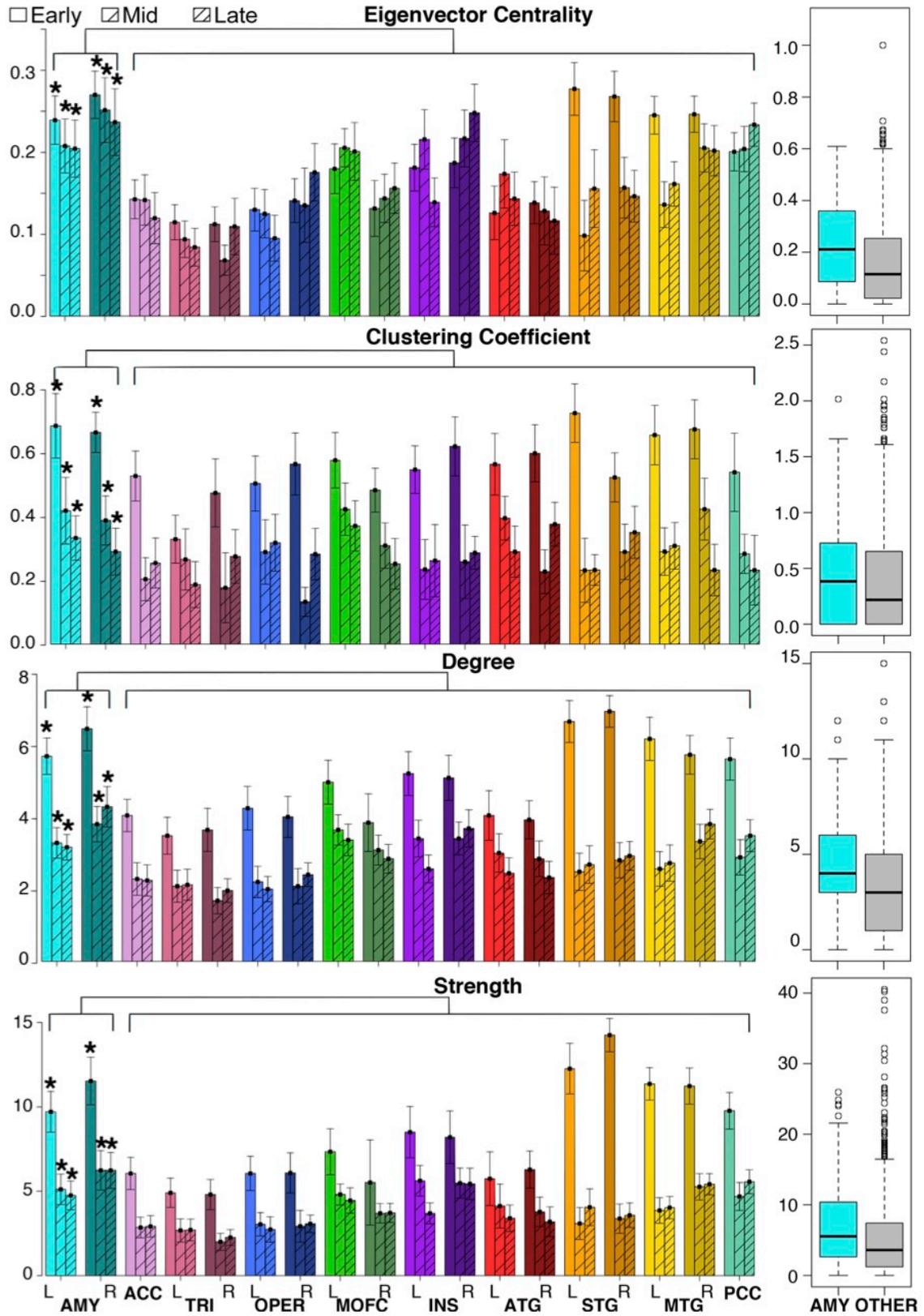

**Figure 3** . *Graph-theoretic estimates of PNN topology and amygdala centrality*

(a) *Left:* Contrasting right and left amygdala (far left) with other PNN nodes. Both left and right amygdala had higher *eigenvalue centrality* across all three time windows ($F_{1,1164}=28.80,p<0.0001$). *Right:* boxplot contrasting amygdala and all other regions combined. (b-d) Likewise, the amygdala displayed higher levels of *clustering coefficient* ($F_{1,1164}=5.48,p<0.0194$), *degree* ($F_{1,1164}=20.34,p<0.0001$), and *strength* ($F_{1,1164}=20.28,p<0.0001$) than averages of other PNN nodes across all time windows. All p values are FDR corrected.

### *Diffusion weighted estimates of PNN white-matter fiber pathway connectivity*

The graph theoretic model of functional connectivity we developed thus far is unconstrained in terms of structural pathways by which such communication might be effected. We therefore calculated the patterns of probabilistic tractography between these functional PNN nodes, the group-averaged results of which are displayed in *Fig. 5* (see *Fig. S3* for an expanded view).

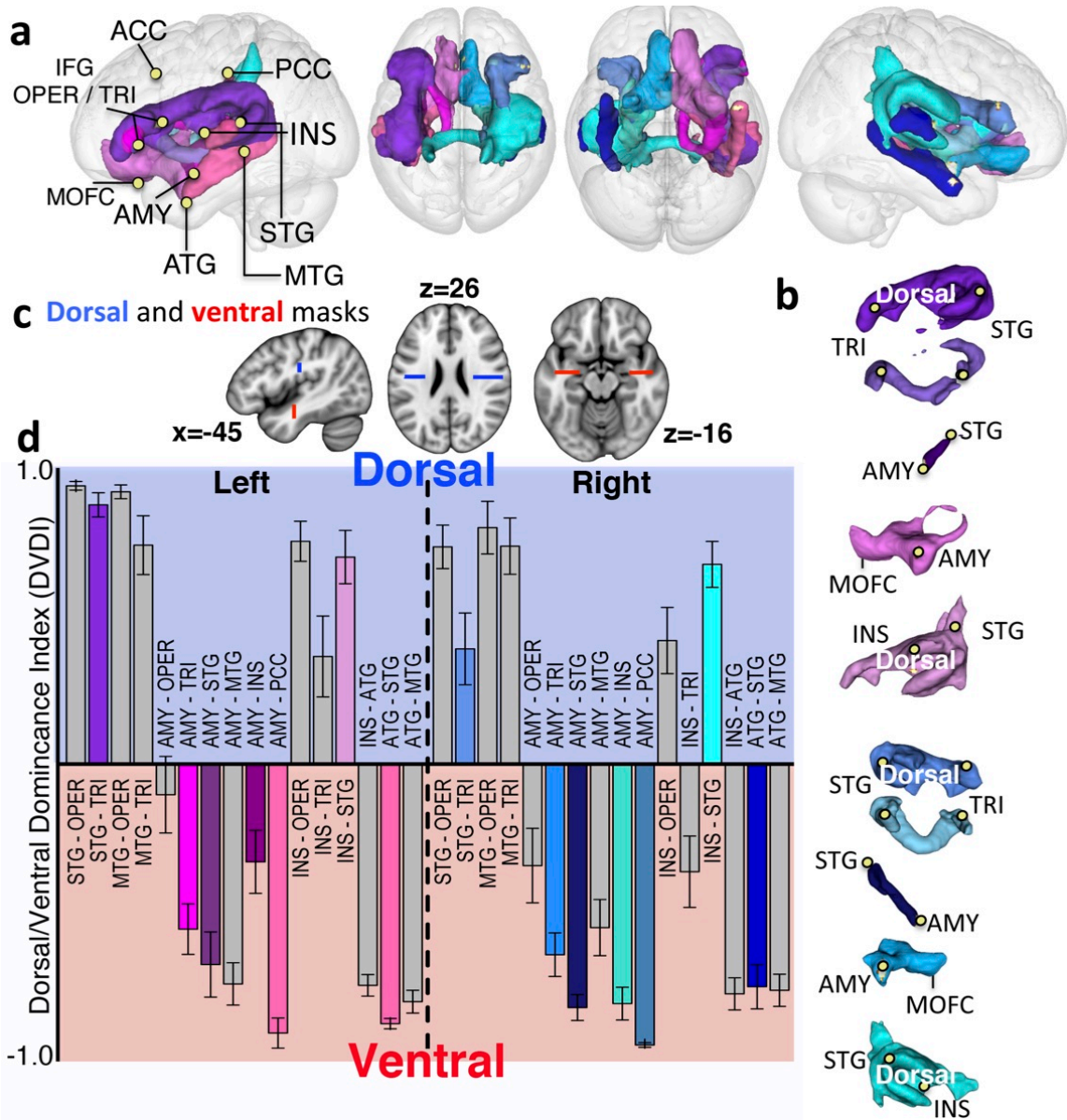

**Figure 5. Probabilistic tractography pathways**

(a) Group-averaged patterns of tractography between functionally and anatomically defined PNN nodes. (b) Major dorsal and ventral pathways interconnecting these nodes, displayed individually from a left and right viewpoint. (c) Waypoint masks bisecting white-matter fiber bundles comprising the dorsal (SLF/AF) and ventral (within the ILF) pathways. (d) Probabilistic tract projections interconnecting ROIs, as contrasted by the waypoint masks. The DVPD indices indicate a high ventral pathway dominance for AMY connections to STG, MTG, INS, PCC, TRI, and right (though not left) OPER.

Our MEG data establish that the amygdala connects with many PNN nodes, including STG, MTG, medial orbitofrontal cortex (MOFC), insula, TRI, and OPER. By creating waypoint masks bisecting the white-matter fiber bundles that comprise the dorsal (SLF/AF) and ventral (EmC) pathway (*Fig. 5C*), we were able to contrast the relative robustness of probabilistic tract projections along the pathways that connect these nodes and their cortical counterparts (thus forming a *Dorsal-versus-Ventral Pathway Dominance index*, DVPDI). As shown in *Fig. 5D*, amygdala connections to STG, MTG, insula, posterior cingulate precuneus (PCC), TRI, and right (but not left) OPER had high ventral pathway dominance. The amygdala also robustly interacted with MOFC along the uncinate fasciculus. Importantly, close inspection of the ventral pathway tractography trace (where the EmC is most visible) and the bridge point between the temporal and frontal lobes indicates this interconnection follows fiber projections passing through the external capsule (ExtC) as well as the EmC (*Fig. S6*).

Our tractography also indicates that the insulae predominately connect with the posterior STG PNN nodes via dorsal SLF/AF fibers. However, insula connections with anterior portions of the temporal cortex (ATG) are predominately ventral. Insula communication with IFG is more varied: DVPD indices of insula-OPER tractogaphy indicate a bilateral dominance of the dorsal pathway, but on the other hand, insula-TRI tractography is dorsally dominant in the left hemisphere (LH), whereas ventral pathways dominate in the right hemisphere (RH). Nonetheless, we observed no corresponding left-versus-right hemisphere insula-TRI *functional* connectivity differences ($t_{139}$=-1.29, p=0.20).

Finally, tractographic connections between ATG to both STG and MTG exhibited a high ventral pathway dominance. Conversely, OPER and TRI tractographic connections with STG and MTG were predominately dorsal. One exception was RH STG-TRI, which displayed significantly reduced dominance compared to its LH counterpart. This lower RH STG-TRI dominance was in part due to the fact that DVPD index variability across subjects was higher in the right compared to the left hemisphere (see *Fig. 4D*).

**DISCUSSION**

Here we define a neural network that orchestrates prosodic processing, that we term the Prosody Neural Network (PNN), revealing critical participation by the amygdala and the insula not previously known to be a part of this system.

Millisecond-resolution MEG enabled us to characterize orchestrated network activity patterns and their temporal evolution, while MR Diffusion Imaging and tractography between connected PNN nodes enabled us to identify the paths that could make such functional connections. Functionally, graph theoretic metrics of functional network connectivity indicated that the amygdala maintains robust and sustained connections with several cortical PNN nodes, supporting our contention that it acts as a central hub for prosodic processing. As such, the amygdala could modulate activity within the network, or integrate information passing through the nodes it contains. Structurally, our tractographic analysis indicates that the amygdala has direct interactions with STG/MTG and IFG via ventral white-matter projections, in which fibers passing through the EmC connect temporal and frontal cortices (*Fig. S5; see also footnote*[†]). Our results also confirm that the amygdala robustly connects with insula, which, in turn, maintains interactions with both temporal and frontal PNN nodes by projecting onto STG and IFG-OPER via dorsal pathway arcuate fibers. *Fig. 5* illustrates the combined functional-structural PNN model resulting from our integrated neuroimaging approach.

---

[†] Parsing fiber bundles that interconnect STG, TRI, and OPER nodes in IFG has proven difficult. Despite indications that principal fibers pass through EmC (Saur et al. 2008), it is not clear that these are completely distinct in origin from those traversing ExtC (Dick and Tremblay 2012). In fact, ExtC fibers connecting the amygdala to STG lie close to the EmC, along with the medial and inferior longitudinal fasciculi (ILF) that putatively project to MOFC and perhaps IFG (Makris and Pandya 2009). These findings bolster the possibility that a ventral path connecting the STG, amygdala, and IFG has both ExtC and EmC fiber components.

Two major conclusions emerge from this model. First, the timing, pattern, and structural paths of amygdala-cortical functional interactions we show to occur during prosodic perception indicate key limbic contributions to vocal communication. This key limbic contribution comprises both extensive functional and structural connectivity to central regions of the PNN, and which has yet not been demonstrated in former studies (T Ethofer et al. 2012; S Frühholz and Grandjean 2012; S Frühholz et al. 2015; T Ethofer et al. 2006; S Frühholz, Gschwind, and Grandjean 2015; Peron et al. 2016). Our inclusion of the amygdala as a critical hub in speech processing contrasts with current concepts of speech and language that focus on cortical, cognitive, and motor contributions (Friederici 2012; Rauschecker and Scott 2009). In these models, STG/MTG and the TRI and OPER nodes within IFG purportedly act on the product of temporal-cortical processing which integrates features of the speech stream into a coherent percept (Hagoort 2014; Schirmer and Kotz 2006b). Though our results focus on affective speech signals, we contend that *all* interpersonal conversation employs prosody, because communication is intrinsically social (Tomasello 2010); thus, communicative processes are likely to routinely recruit limbic and emotional processing resources.

Applying the same experimental paradigm used here, we previously found that the amygdala and insula display correlated increases in activity that parallel rising emotional salience of acoustic features, especially those having levels that strongly predict emotion identification (Leitman et al. 2010; Leitman et al. 2011). These findings, along with our amygdala-STG/MTG functional connectivity and tractography results, which point to a ventral pathway for this communication, all indicate that amygdala processing collaborates with STG/MTG to form an emotional percept from the incoming speech signal. Monkey studies further indicate that the amygdala may rapidly (within 20 ms) process auditory information originating from the thalamus or perform primitive cortical processing to generate social orienting responses (Armony and LeDoux 2010). Thus, it is likely that amygdala-cortical interactions involved in audio-affective semantic processing take the form of a recurrent loop consisting of multiple exchanges in which affective and semantic relevance becomes increasingly more sophisticated (Schirmer and Kotz 2006a; Sascha Frühholz, Trost, and Kotz 2016). In these exchanges, orchestrated interactions with the MOFC may facilitate the elaborated appraisal (Sascha Frühholz, Trost, and Kotz 2016) and re-experiencing–or embodiment of the affective signal thought to be pivotal in comprehending the social intention of others (Niedenthal 2007). As we continue this line of research, we plan to examine this amygdala-based loop by measuring the spectral dynamics of the coupling patterns between PNN nodes, which will enable us to create directed or causal network models of PNN function.

The second conclusion that emerges from our PNN model regards the dual role of insula in both affective and motoric processing, and its clarification of the bifurcated processing streams anchored by STG and IFG (TRI and OPER) nodes (*Fig. 6*). We began this study to determine if the amygdala contributes to a ventral pathway connecting STG and IFG, whose role is to extract and evaluate semantic meaning from the auditory stream (Rauschecker and Scott 2009; S Frühholz and Grandjean 2013). However, our data also indicated that the amygdala robustly connects with the insula, which in turn projects to STG and IFG via the dorsal AF stream. These dorsal insular contributions to prosodic processing are intriguing because their connections to basal ganglia and the supplementary motor area are implicated in sensorimotor impairments concerning the preparation, production, and perception of prosody. For example, Parkinsonian dysprosodia is linked to striatal dopamine reductions (Benke, Bösch, and Andree 1998; Caekebeke et al. 1991). At the same time, insula-striatal communication may also reflect emotional dysprosodia antecedents, given the strong limbic input that the    striatum    receives    (Pichon    and    Kell    2013).

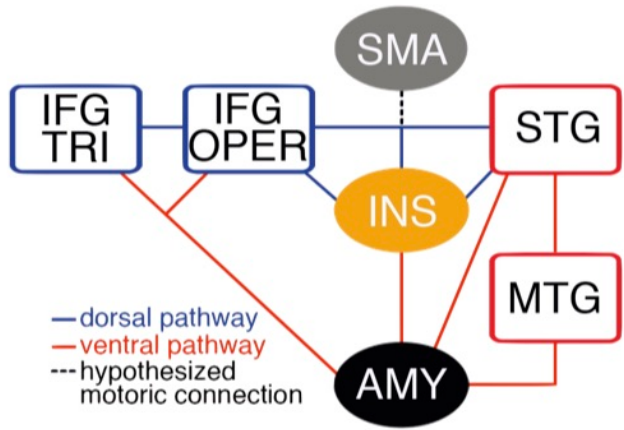

**Figure 6 Temporal-frontal cortical pathways of the PNN that are active during prosodic communication**

This model shows connections between prosody regions of interest indicated by our findings. Functionally, the red connections are involved in extracting and integrating acoustic features to form an emotional prosodic percept; processing within these regions increases as the emotional salience of acoustic features becomes more enriched. Blue lines indicate connections that have increased activity when prosodic stimuli contain low salience and are emotionally ambiguous. Anatomically, the upper blue line depicts the dorsal pathway comprising portions of the SLF and AF connecting PNN temporal cortical structures (STG, MTG, OPER, and TRI). The red lines comprise the ventral pathway, which utilizes fiber bundles running medially and transecting dorsal aspects of the basal ganglia through the ExtC and possibly the EmC.

Our results are consistent with evidence that the dorsal and ventral pathways can act synergistically to process speech syntax and morphology (Rolheiser, Stamatakis, and Tyler 2011). They also harmonize with hypotheses like those produced by Hickok and Poeppel (Hickok and Poeppel 2007), in which motor representations of speech can be integrated with ventrally-based auditory representations during language perception tasks. With these facts in mind, our data support a model in which: a) amygdala-insula connectivity vertically integrates ventral pathway processing that combines auditory features into an emotional percept, and b) relays the results to the dorsal pathway that contributes motor representations of prosodic articulation and speech planning.

A limitation of our approach is that reliance on MEG for dynamic connectivity estimates prevented us from reliably obtaining activity estimates from supplementary motor areas (SMA) (see methods). To rectify this, in future studies, we will utilize combined EEG-MEG source modeling and explicit Diffusion Weighted Imaging mapping of insula-SMA projections to more precisely map insula contributions to prosodic processing and its putative role in vertically integrating ventral and dorsal pathways.

In conclusion, as E. Colin Cherry remarked nearly a half century ago, "communicatory signals are not sent or received, they are shared" (Cherry 1978). Speech and language, even when not explicitly emotional, are nevertheless intrinsically social in nature. Our findings highlight the important contributions that the amygdala makes to prosodic processing, and perhaps to more general aspects of vocal communication. By delineating influences of the amygdala in prosodic functioning, we suggest that the classic Wernicke–Lichtheim–Geschwind model of speech and language be extended to incorporate ventral pathway and limbic processing contributions.

**Materials and Methods**

***Participants.*** Twenty-eight healthy adults (22 males; 26.14±7.37 years of age; 14.92±1.76 years of education) with no history of mental illness participated in the study. Two subjects were removed entirely because they produced poor quality data. Of remaining subjects, one produced poor quality MEG data, three did not complete the diffusion weighted imaging scanning, and probabilistic tractography procedures failed on an additional subject. Thus, the final MEG data presented here contained 25 subjects, and final tractography data contained 22 subjects. A five-minute acoustical test established that subjects had no history of hearing loss. All were right-handed. Informed consent was obtained, and study methods were approved by the University of Pennsylvania Institutional Review Board.

***Prosody Task***. Our assessment of prosody performance relied on the paradigm we employed in our previous fMRI studies(Leitman et al. 2010; Leitman et al. 2011), which involved a subset[‡] (N=26: 8 happy, 8 fear, and 10 anger stimuli) of the stimuli described by Juslin and Laukka (Juslin and Laukka 2001). That study featured British English speakers expressing semantically neutral sentences (*e.g.*, "it is 11 o'clock") while projecting *happiness*, *anger* or *fearful intent*. For the experiments described below, each emotion was presented 15 times in the electroencephalographic-magnetoencephalography (EEG-MEG) trials and twice more in the subsequent fMRI tests. Participants were asked to choose which emotion was represented, given four choices (*i.e.*, the three intended emotions, or no expression). In a separate set of experiments, we determined relationships between prosodic features, identification accuracy, and certain functional and structural brain metrics. It is notable that the overall group prosodic identification accuracy during MEG recording was 63±7%, roughly 2.5 times greater than chance and consistent with our previous research (Leitman et al. 2010; Leitman et al. 2011). Finally, although both EEG and MEG data were jointly collected for most subjects, the analysis presented here solely utilized MEG sensor data. We intend to contrast the MEG and combined EEG-MEG source models in a separate study.

**Data Collection and Analysis**. To demonstrate that the amygdala performs as a principal PNN hub, we integrated our prosodic identification paradigm with MEG. By applying fMRI weighted-source modeling to the millisecond-resolution MEG time series, we identified pairwise cross-trial functional connectivities between PNN nodes. Graph theoretic topological analyses were also used to quantify the extent to which the amygdala acts as a hub for PNN prosodic processing.

---

[‡] The duration of the experiment was of critical concern for two reasons: first, this study was part of a larger research program contrasting a healthy and a clinical population with known cognitive and attentional impairments. Second, our MEG hardware setup made recording a single file larger than 2GB problematic, as did joining separate MEG recordings. To shorten the study, we opted to reduce the number of stimuli to a minimum. Prior evaluations of the correlation between the emotional cue salience for happiness (increasing fundamental frequency variability) and fear (decreasing fundamental frequency variability) reveal that the two omitted stimuli for both fear and happiness did not occupy wholly unique points along the cue X detection probability regression line. Consequently, in the interest of time these stimuli were omitted.

Magnetic Resonance High Angular Resolution Diffusion Imaging (HARDI) and probabilistic tractography were used to determine if ventral projections between the amygdala and STG were more robust than their dorsal alternatives. *Fig. 1* depicts the overall data collection and analysis path.

*Magnetoencephalography.* MEG data were recorded in an actively shielded room using an Elekta Neuromag whole-head system (Elekta Neuromag® Vectorview[TM] Helsinki, Finland) consisting of 306 channels (102 magnetometers and 204 planar gradiometers). Subject placement relative to the MEG sensors was monitored continuously via four head-position indicator coils attached to the scalp. Bipolar channels above and below the eye and others located at the right and left clavicles, were employed to measure eye blinks (EOG) and cardiac activity (ECG), respectively. Electrode impedances were maintained below $10^4$ ohms, and we sampled the analog data from 0.4 to 330 Hz online at 1000 fs intervals. The 1000 fs-sampled data were processed with an Elekta Temporal Signal Space Separation (TSSS) filter (Taulu and Simola 2006), and then down-sampled to 500fs after conditioning with a 0.8-100 Hz bandpass filter. Processed MEG data were co-registered with corresponding T1 structural MRI images (described below), which provided the input for the Freesurfer distributed source model (http://surfer.nmr.mgh.harvard.edu/) surface reconstructions; the standard pial and white-matter surfaces that Freesurfer created were augmented by subcortical surfaces of the hippocampus and amygdala (see below).

These surfaces were generated from Freesurfer subcortical segmentation, and then merged to the cortical surfaces so they could be incorporated into the Minimum Norm Estimate (MNE) distributed source model software via custom scripts created within MNE and MNE-Python libraries (Gramfort et al. 2014). We also constructed structural masks for our *a priori* Regions of Interest (ROIs; *i.e.,* STG, IFG), using the APARC 2009a atlas in Freesurfer together with the Harvard-Oxford atlas in FSL (Smith et al. 2004). MEG data were processed to remove large spikes (magnetometers <4.0 *pT*, gradiometers <4.0 *mT*, EOG <150 *µV*), and EOG and ECG artifacts were corrected using Independent Components Analysis (Gramfort et al. 2014), which most often removed the 1st and 2nd components that likely reflected these artifacts.

Trial epochs were constructed to reflect -0.4 to 1.9 seconds Post-Stimulus Onset (PSO). Bad channels were defined as those that contained artifact rejection/corrections of 15% or more of all trial epochs. Using MNE software, we used the average of all trials for the L2 minimum-norm model (Hämäläinen et al. 1993), thereby generating an anatomical surface-based source model of sensor activity. To spatially restrict this solution, we overlaid a z-map reflecting blood-oxygen-level dependent (BOLD) activation in the analogous fMRI prosody task, so we could choose, within the structural ROIs (as restricted by this fMRI z-map; Z>2.95), the MEG vertex that reflected "center of mass of source activity." This source was calculated using the Dynamical Statistical Parametric Mapping (dSPM) inverse operator, which provides a signal-to-noise estimate of activity that is optimal for deep sources and facilitates fMRI-MEG integration. We then grew 4-mm spheres centered on this vertex, thus identifying a source for each individual subject and each functional-structural ROI. These ROIs are listed in *Table 1*. For every trial, the standard MNE inverse operator was used to identify activations within the individualized sources.

To reduce these activations into a single time series, we applied a Singular Value Decomposition to the time courses within the spheres using the scaled and sign-flipped first right-singular vector as the sphere time course. This approach yielded a single time series vector per trial for each ROI. Matrices containing this processed trial data for each *a priori* ROI were exported for time-frequency and connectivity analysis using custom Python/Matlab routines, which can examine the temporal-spectral connectivity patterns among ROIs using graph-theoretic measures of network characteristics (as described below).

***Subcortical Sources: Amygdala.*** Our previous fMRI data identified multiple subcortical regions that are activated in response to modulations of prosody cue saliency, including amygdala and insula (Leitman et al. 2010; Leitman et al. 2011). We hypothesized that the amygdala plays a pivotal role in social orienting by prioritizing incoming stimuli for approach-avoidance decisions(Pessoa and Adolphs 2011; Davidson 2000). To test this, we estimated MEG source activities within this subcortical region (although it is a deep structure, recent publications have established the feasibility of detecting MEG signals from the amygdala(Salvadore et al. 2010; Cornwell et al. 2008; Cornwell et al. 2007; Luo et al. 2007; Hung, Smith, and Taylor 2012; Mišić et al. 2016)) Following an approach analogous to that outlined by Balderston *et al.* (Balderston et al. 2013),        and        also        implemented        in        *Brainstorm*        software (http://neuroimage.usc.edu/brainstorm/), we added surface renderings of the amygdala and hippocampal complexes to our Freesurfer-generated pial cortical surface. We populated this additional surface space with perpendicularly-oriented virtual source dipoles (similar to those found on the cortical surface). Merging the hippocampus with the amygdala helped ensure proper alignment to the cortical surface. To adjudicate whether amygdala signals were distinct from those of the adjacent cortical surface, we contrasted the averaged time-course and spectral profile of signals from both amygdalae with the insula, which is their nearest cortical surface neighbor. As seen in *Fig. S1b*, although amygdala and insula signals are similar in time course, they are nevertheless clearly distinct in both evoked activity time pattern and their respective temporal-spectral profiles.

   ***ROI-ROI Connectivity and the Prosody Neural Network.*** The goal of our processing stream was to produce a graphical representation of the communication patterns between functionally active PNN regions using time series of MEG sources. To accomplish this, trial time series data reflecting brain responses to all prosodic stimuli were split into three equal samples containing *early* (0-600 ms), *middle* (600-1200 ms), and *late* (1200-1800 ms) windows of PSO activity. Data for these three periods were then spectrally decomposed using a multi-taper method with Digital Prolate Spheroidal Sequence (DPSS) windows for frequencies ranging from 2-40 hz. Across this frequency range, the magnitude of between-node connectivity was measured by testing the oscillatory phase synchronization between the 18 PPN nodes in a pairwise manner. Specifically, we employed a weighted and unbiased estimate of the Phase Lag Index (wPLI)(Vinck et al. 2011) which measures the instantaneous, or zero-phase, connectivity between sources. wPLI tests the statistical interdependencies of each nodal time series on its counterpart by quantifying the magnitude of node-to-node oscillatory phase distribution asymmetry differences(Stam, Nolte, and Daffertshofer 2007). wPLI was chosen over other connectivity metrics (*e.g.*, imaginary coherence) because it is less susceptible to spurious findings of zero-phase cross regional connectivity that are the product of volume conduction and reflective of a single common source(Stam, Nolte, and Daffertshofer 2007). The wPLI ranges from 0 (signifying no connectivity) to 1 (signifying perfect connectivity). wPLI coefficients were subjected to a Fisher transform to normalize their distribution, and allow us to incorporate them into linear statistical models that examine whether wPLI node-to-node connectivity changes over time. For every subject, these procedures produced three 18×18 matrices reflecting connectivity between each pair of PNN nodes during early, middle, and late periods of prosodic processing.

For graphical analysis, we created thresholds for subject matrices by z-normalizing all node-to-node raw maximum connectivity values across all three time windows. This procedure provided an individualized rank (Z-value) for each *'edge'* or node-to-node connection, and a threshold of Z≥0.5 (69[th] percentile) was applied to the wPLI matrix. We employed a Concordance-at-the-Top (CAT) analysis to examine the homogeneity of subject wPLI matrices across the group. This analysis also provided a rationale for establishing our wPLI connectivity threshold (see *Fig. S3* for details), which enabled us to topologically overlay graphical representations of the thresholded wPLI data onto brain anatomy for visualization (*BrainNet Viewer*: http://www.nitrc.org/projects/bnv/ (Xia, Wang, and He 2013)). We also calculated graph-theoretic measures of topological similarity between and across PNN nodes using the *Brain Connectivity Toolbox*[48]. To test our hypothesis that the amygdala functions as a principal hub for PNN processing, we calculated the following dependent variables for each PNN node: 1) *Degree:* the number of edges; 2) *Strength:* the summed weight (wPLI magnitude) of all node edges; 3) *Clustering Coefficient:* the fraction of triangles around a node, which is the fraction of the node's neighbors that are neighbors of each other; and 4) *Eigenvector Centrality:* a relativistic measure of nodal influence predicated on ranking nodes within network connections, based on whether they connect with other highly connected nodes[48]. We then statistically tested whether amygdala values of *degree*, *strength*, *clustering*, and *centrality* exceeded those of the PNN network as a whole using a series of Three-Way ANOVAs (region (amygdala vs. others) x time period x hemisphere), holding an overall Type1 error rate to α<0.05 and adjusting for multiple comparisons using False Discovery Rate (FDR) correction using the method of Benjamini & Hochberg[49].

***Magnetic Resonance Imaging: Acquisition and Analysis.*** We used a 3T Siemens Verio Scanner (Siemens, Erlangen Germany) to acquire the following brain images: (1) <u>T1 structural image</u>: this was based on a T1 MPRAGE (repetition time/echo time = 1900/2.87 ms, field of view= 256 mm, matrix = 256 × 256, slice thickness = 1 mm in 176 slices) volume resolution of 1 mm$^3$. This structural scan was used for anatomic/brain surface reconstruction and the co-registration of EEG-MEG and fMRI data using Freesurfer and FSL software. <u>(2) Diffusion imaging and tractography</u>: Whole-brain 2 × 2 × 2mm isotropic HARDI data were collected in the axial plane using a modified Stejskal Tanner sequence with a spin-echo echo-planar imaging readout (63 diffusion-encoding directions, b-value = 3000 s/mm$^2$, as well as a single unweighted b=0 s/mm$^2$ image at 3T, TR=14,800 ms, TE=111 ms, field of view=256 mm, matrix=128 × 128).

The following analysis procedures were performed: (1) *Probabilistic tractography analysis*: we utilized the Diffusion Toolbox (FDT) from the Oxford Centre for Functional MRI of the Brain (FMRIB). For each subject, HARDI data were pre-processed to correct for eddy current and motion distortions and to model the local diffusion parameters. Crossing fibers were modeled through Markov chain-Monte-Carlo sampling (1000 iterations) of the diffusion parameters of each sampled voxel, using 2 fibers per voxel in accord with Behrens *et al.*[50]. Following this processing, we designated pairs of spherical function-structure ROIs (derived from the fMRI-weighted MEG source analysis) as seeds and termini, and created probabilistic fiber tracts between ROI pairs (using the FSL default parameters of 5000 samples through the probability distributions on principal fiber direction, that tolerated a curvature not exceeding ± 80 degrees, with termination occurring at a maximum of 2000 0.5-mm steps; fibers with a volume threshold of less than 0.01 were discarded). These probabilistic tracks were then linearly warped to a standardized MNI 2×2×2 template brain.

(2) *Functional-Structural Data Integration*:    Probabilistic tracks reflecting connections between PNN nodes were considered significant if: a) they were present in the majority of subjects, and b) displayed time series functional connectivity that exceeded Z=0.5 (or 69th percentile) of all possible node-node connectivity estimates across all time windows. These tracks were then interconnected in the manner that best conformed to the patterns provided by weighted undirected graphs of the ROI time series.

*(3) Estimating Dorsal Versus Ventral Pathway Dominance in Amygdala-Cortical Projections.* We assessed the principal route used by the amygdala to communicate with cortical PNN nodes by contrasting the robustness of projections between amygdala and PNN nodes located in temporal and frontal cortex, and passing through ventral versus dorsal waypoint masks situated along these pathways (following the approach outlined by Frühholz *et al.*[21]). Path projection robustness was operationalized from the 'waypoint totals', as the total number (max=5000) of tracks generated from the seed node (amygdala) mask that reached at least one of the voxels in the terminal mask, (*i.e.,* STG, MTG, or IFG-triangularis (TRI) and IFG-opercularis (OPER)) via the waypoint (dorsal and not ventral, or ventral and not dorsal) mask and was not rejected by inclusion/exclusion mask criteria (described above; also see http://fsl.fmrib.ox.ac.uk/fsl/fslwiki/FDT). Statistical testing was conducted in R (r-project.org) and utilized the following libraries: Hmisc, multcomp, nlme, plotrix, plyr, psych, and reshape2.

### *Acknowledgments*

The authors would like to thank Drs. Wolf, Janata, Henderson, and Luther for reviewing and editorial suggestions. The authors would like to thank Dr. Russell Shinohara for his advice and expertise with the statistical procedures employed in this study. The authors would also like to thank Dr. Christian Kohler, Jamie Rundino, John Dell, Rachel Golembski, Charlie Fisk, and Peter Lam (Children's Hospital of Philadelphia), who assisted in the supervision and collection of the MEG/MRI data, as well as Margo Wilms (University of Groningen) and Steven Meisler (University of Pennsylvania) for their contributions to the data analysis.

This work was supported by the National Institute of Mental Health (K01-094689 to D.I Leitman) and a NARSAD Young Investigator Award from the Brain and Behavior Research Foundation (D I Leitman).

**Author Information and competing interests**

D.L., J.B., J.C.E., and T.P.L.R. designed the experiments. D.L. and K.G. collected and analyzed the data. D.L. prepared the initial manuscript, and all authors contributed to further edits of the paper.

The authors declare no competing financial interests.

# Supplementary Information for

## Amygdala And Insula Contributions To Dorsal-Ventral Pathway Integration In The Prosodic Neural Network

David I. Leitman[1], Christopher Edgar[2], Jeffery Berman[2], J. Krystal Gamez[1,3],

and Timothy P.L. Roberts[2]

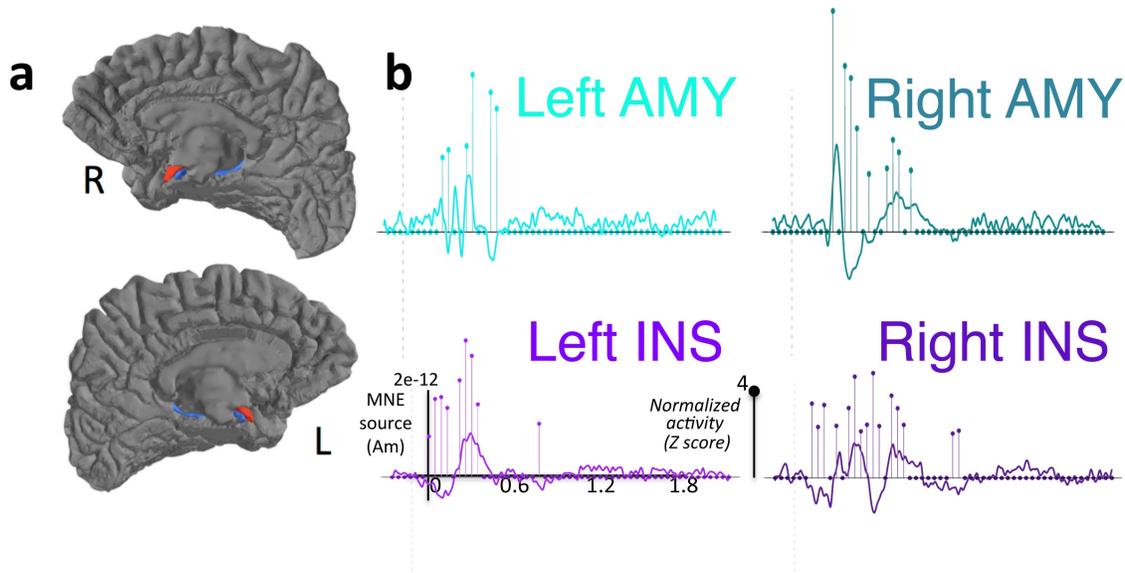

***Figure 1.*** **MEG Source Activity from Amygdala.**

(a). Example of a surface model that incorporates both amygdalae (red) and hippocampus (blue) renderings. As with the cortex, our MNE estimates include dipoles oriented to amygdalae surfaces. (b). Grand-averaged group waveforms of task event-related activity for amygdala and insula, including normalized within-ROI activity summaries (stem plot) as an overlay. The insula is the closest cortical surface point we have to compare with the amygdala, and their time courses are clearly distinct.

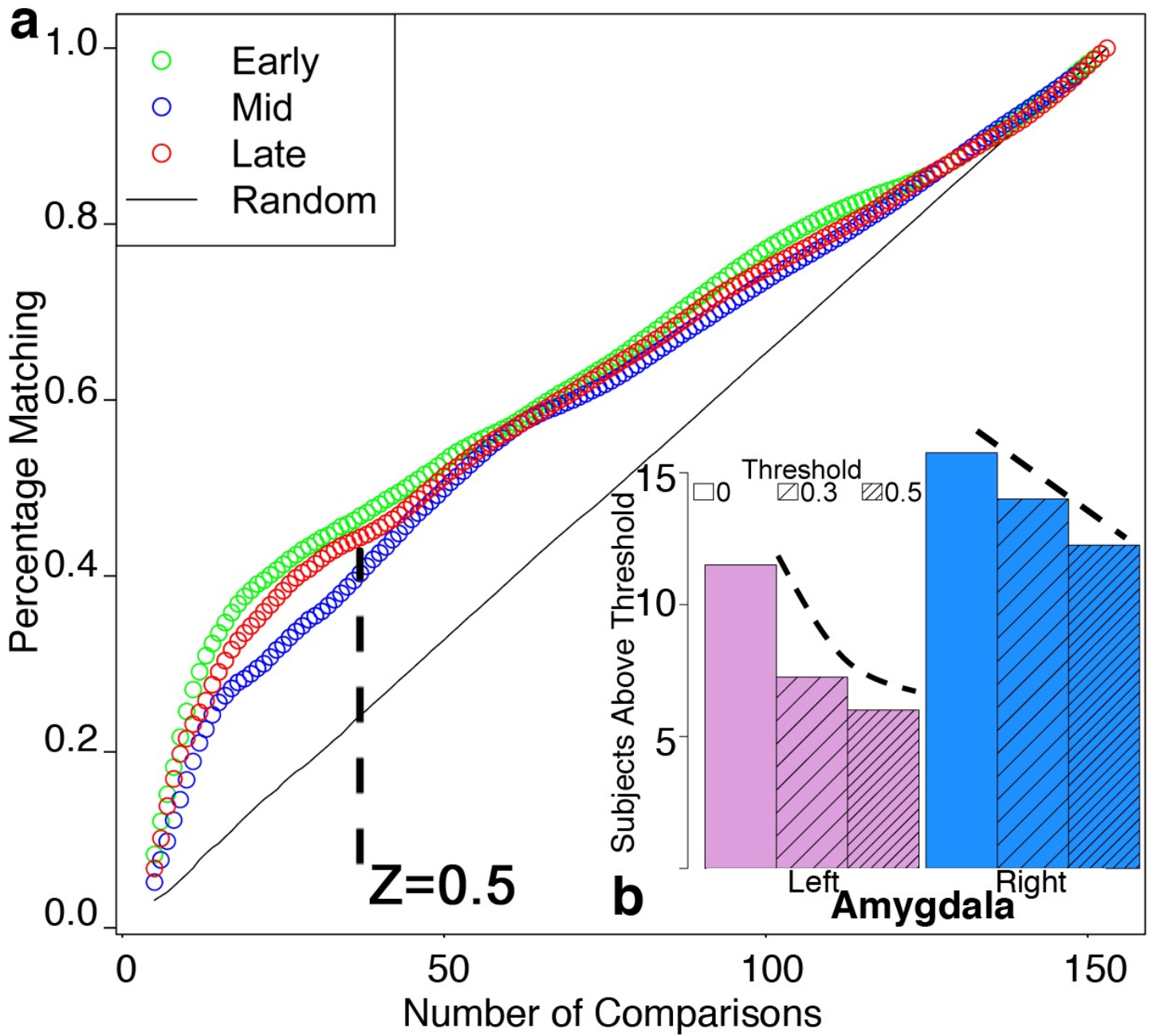

**Supplementary Figure 2. Estimation of Homogeneity in Connectivity Patterns Across Subjects Using Concordance-At-The-Top (CAT).**

(a). We measured subject sample homogeneity by iteratively and randomly assigning them to either of two subsets (10,000 repetitions: 12 in the first group, 13 in the second), and testing the concordance between each sample's top (*i.e.,* highest ranked) edges, across incrementally larger numbers of "top" edges. The green, blue, and red traces represent the concordance rates (y-axis) as a function of the "top" size (x-axis) for the normalized (non-threshold) wPLI values from early, middle, and late time windows, respectively. The line running semi-parallel to these traces represents random chance, and the intersecting dashed line indicates where the connectivity threshold of $Z=0.5$ lies; this line roughly corresponds to the change in slope of the CAT plot for the early window shown in green. (b). Number of subjects whose amygdala-cortical (AMY-STG, AMY-INS, AMY-TRI, and AMY-OPER) normalized wPLI connectivity magnitude exceeded the $Z=0.5$ threshold, along with two more liberal cutoffs of $Z=0$ and $Z=0.3$. The increase in supra-threshold subjects that parallels decreases in connectivity thresholds indicates that the functional connectivity patterns are not idiosyncratic to our selected threshold.

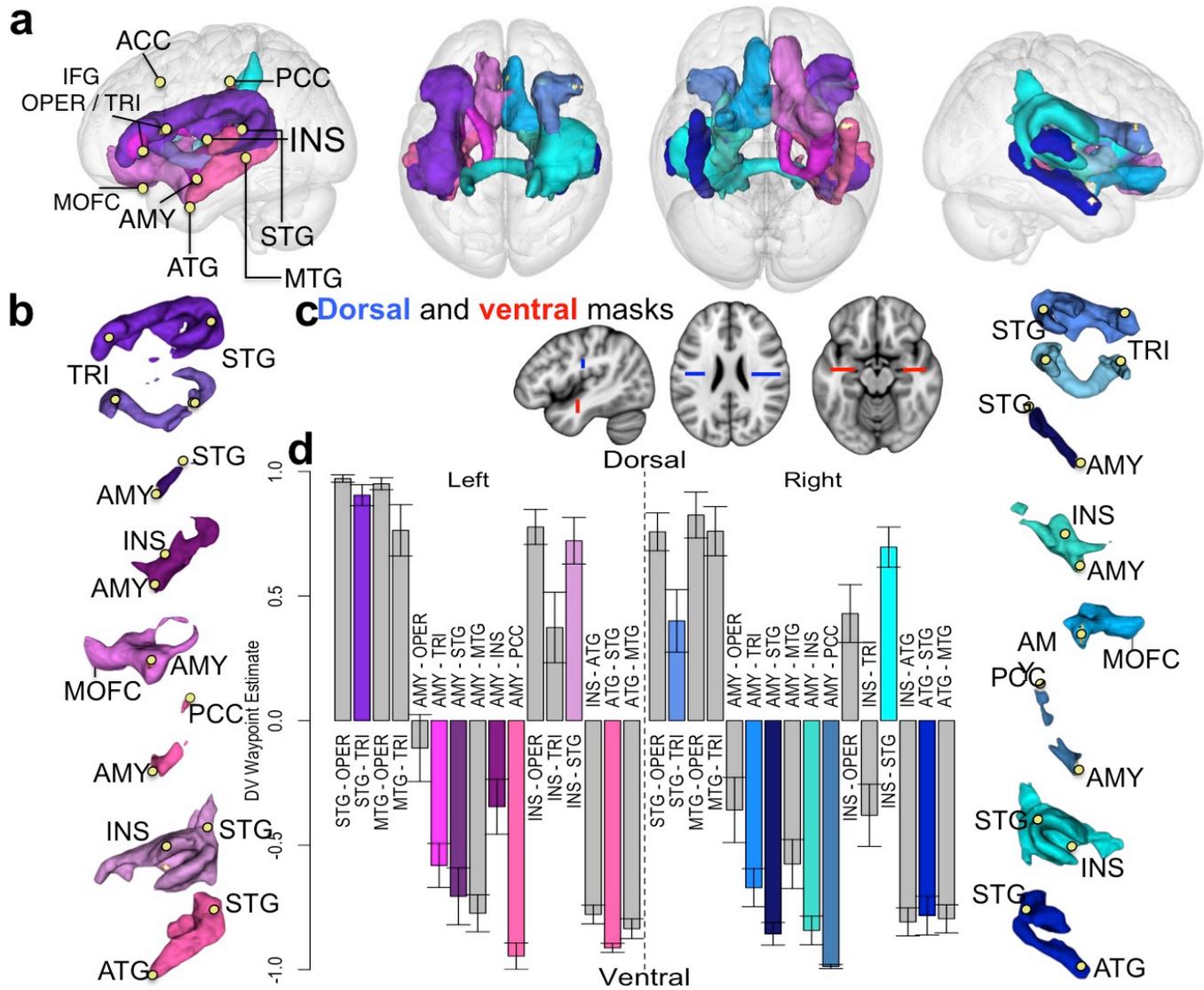

**Supplementary Figure 3:** Expanded Figure 5 - Probabilistic Tractography Pathways

This figure is an expanded view showing tractography of suprathreshold functional connectivity estimates, and includes PCC to AMY pathways.

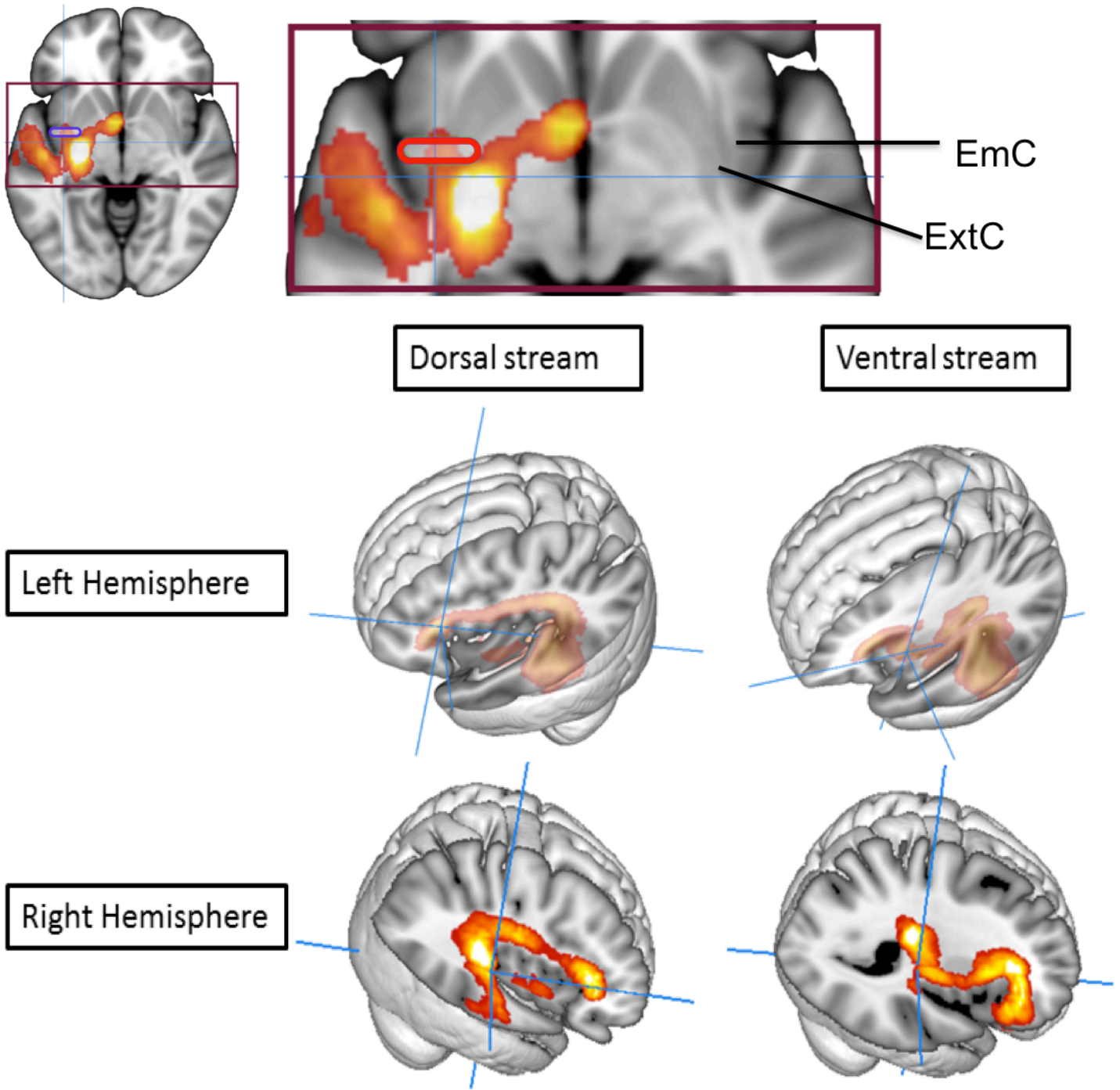

**Supplementary Figure 4: HARDI Probabilistic Fiber Close-up and Alternate Views** (a) This close-up highlights the extreme and external capsules, with the ventral path we established overlaid in red. The red rectangle highlights the ventral EmC mask region we used for ventral path tractography. In these images, it is difficult to localize the ventral pathway as part of the EmC, especially given the proximity of the ExtC and the resolution of our techniques (see footnote in discussion section for details). (b) Three-dimensional renderings of dorsal and ventral pathways at the point at which they bridge temporal and frontal cortex

| Frequency Band (Hz) | M±SD |
|---|---|
| 2 | 0.011±0.0088 |
| 4-9 | 0.021±0.011 |
| 10-16 | 0.027±0.011 |
| 17-30 | 0.017±0.0047 |
| 31-40 | 0.011±0.0040 |

*Supplementary Table 2. Peak wPLI Connectivity Within Each Frequency Band*